\documentclass[prf,onecolumn,english,floatfix,longbibliography]{revtex4-1}
\usepackage{blindtext}
\usepackage{graphicx}
\usepackage[utf8]{inputenc}

\usepackage{epstopdf, epsfig}
\usepackage{amsmath}
\usepackage{float}
\usepackage{stmaryrd}
\usepackage{braket}
\usepackage{xcolor}

\usepackage{bm}

\newcommand {\tck}{\textcolor{black}}

\begin{document}
\title{Subcritical transition to turbulence triggered by a magnetic dynamo}
\author{F.Daniel $^1$}
\email{florentin.daniel@phys.ens.fr}
\author{L.Petitdemange$^2$}
\email{ludovic.petitdemange@upmc.fr}
\author{C. Gissinger$^{1,3}$}
\email{christophe.gissinger@phys.ens.fr}

\affiliation{
$^1$ Laboratoire de Physique de l'Ecole normale superieure, ENS, Universite PSL, CNRS, Sorbonne Universite, Universite de Paris, Paris, France\\
$^2$ Laboratoire d'Etudes du Rayonnement et de la Matière en
Astrophysique et Atmosphères (LERMA), Observatoire de Paris, PSL,
CNRS, Sorbonne Université, Paris, France\\
$^3$ Institut Universitaire de France (IUF), Paris, France}

\begin{abstract}

It has recently been shown that a significant slowdown of many stars can be attributed to the emergence of a strong magnetic field within the radiative region, where heat is transferred through radiation in a stably stratified layer. Here, we describe how this transition can be understood as a subcritical bifurcation to small-scale turbulence in linearly stable flows. The turbulence is sustained by a nonlinear mean-field dynamo and can be observed down to relatively small differential rotation, arbitrarily far from the linear onset of any hydrodynamic instability.  In this regime, turbulent fluctuations provide diffusivity-free transfer of angular momentum that increases the transport generated by the magnetic field triggering the turbulence. Finally, we present a simple nonlinear model that captures this scenario and can be used as a general description of the transition to turbulence in astrophysical flows, as long as it involves a competition between a large-scale dynamo, and a small-scale magnetic instability.
\end{abstract}

\maketitle

\section{introduction}

Turbulence is one of the most ubiquitous phenomena in astrophysics, as the majority of astrophysical bodies involve turbulent flows: convective stars \cite{Brummell1995}, planetary cores \cite{Aurnou2015}, accretion disks, \cite{Balbus1991} or the interstellar medium \cite{Rosner1989} all exhibit turbulent features. But on the other hand, astrophysical flows are sometimes subject to strong stabilizing effects, such as density stratification or global rotation, which prevent, limit, or mediate the transition to turbulence. 

The gas around accretion disks, for example, exhibits a Keplerian rotation known to be linearly stable with respect to the  centrifugal instability~\cite{Rayleigh}. In the absence of such a simple primary instability, it is therefore more difficult to clearly identify a route to turbulence, unlike other rotating systems like Taylor-Couette flows. This situation has given rise to intense theoretical activity aimed at finding an alternative destabilization in these systems, such as baroclinic instability~\cite{Klahr_and_Bodenheimer_2003}, vertical convective instability~\cite{Lesur_et__Ogilvie_2010}, or gravitational instability~\cite{Toomre_1964}, to name a few. Understanding how the turbulence in these disks is generated from one or another of these instabilities remains a long-standing problem in astrophysical fluid dynamics.
A similar situation is encountered in the radiative zones of stars, which are regions where the strong density stratification is such that heat is transported by radiation rather than by turbulent convective motions. These layers generally display differential rotation but the stratification  suppresses thermal convection and strongly delays the generation of shear instabilities \cite{Richardson1920}. As with accretion disks, this seems to contradict the observation of efficient transport of angular momentum or chemical elements in these stably stratified regions, which can only be explained with a more complex flow than simple axisymmetric differential rotation. Several hydrodynamical mechanisms have been proposed to explain this paradox, among which one of the most popular is the Goldreich–Schubert–Fricke (GSF) instability in which the thermal diffusion somehow reduces the stabilizing effect of the  stratification~\cite{Chang2021}  (see \cite{Garaud_2021} for a review of shear instabilities in radiative stars).

An alternative explanation for the destabilization of rotating flows is the role played by the magnetic field when the fluid is electrically conducting. With the so-called magnetorotational instability (MRI), a centrifugally stable flow with a rotation profile decreasing outward (like Keplerian rotation) becomes linearly unstable if the plasma is subjected to a magnetic field \cite{Balbus1998, Ji2023}. Another magnetic destabilization is the Tayler instability, a pinch-type instability in which the toroidal axisymmetric magnetic field generated inside a stellar radiative zone becomes unstable to nonaxisymmetric perturbations when its magnitude is too large~\cite{Tayler1973}.\\

With or without a magnetic field, what happens to these instabilities far from their threshold and how they can lead to turbulence remains a very active open question.  The magnetic scenario is particularly appealing, however, as it is well accepted that the nonlinear stage of MRI can lead to fully developed turbulence~\cite{Guseva2015}. On the other hand, dynamo theory, which describes the process by which a conducting fluid maintains a magnetic field, suggests that turbulence is very efficient in sustaining a magnetic field~\cite{Moffatt2019}.  This complex interplay between turbulence and dynamo has led several authors to postulate that a subcritical transition to turbulence~\cite{Rincon2008, Rincon2019} should be expected when a magnetic dynamo is involved. Over the last few decades, this idea of a transition to turbulence triggered subcritically by the presence of a magnetic field has therefore become very popular, and this scenario has been reported in many different studies~\cite{Riols2013, Herault2011, Hughes1995}.  Among the many other possible route to turbulence, this offers a simple general picture of the transition to turbulence in conducting fluids: an initially stable flow becomes unstable due to a magnetic instability, which amplifies the turbulence through some nonlinear processes. This turbulence then triggers a magnetic dynamo, sustaining the magnetic energy that initially fuels the instability. Such a mechanism is inevitably subcritical, as the generation of both magnetic field and turbulence depend on each other, so that only a disturbance of finite amplitude can trigger this nonlinear amplification loop.\\
 
This search for the origin of magnetism or turbulence is intrinsically linked to the equally important question of finding a mechanism for the transport of heat or chemical elements in astrophysical systems. 
In rotating systems, angular momentum (AM) transport also plays a crucial role: the enormous infall of gas and matter around accretion disks \cite{Balbus1991, Balbus2003}, or the massive slow down of the inner part of radiative stellar layers \cite{Mosser2012, Aerts2019}, are the direct consequence of highly efficient angular momentum transport.  This transport can be achieved by many different processes, such as meridional circulation~\cite{Mathis2013} or the propagation of waves~\cite{Pincon2017}. But the two most convincing sources of AM transport remain the existence of turbulence and/or the generation of a magnetic field, the former providing significant turbulent dissipation while the latter can produce a strong magnetic torque. Naturally, very different predictions can be made for this transport~\cite{Fuller2019, Spruit2002, Chang2021}, depending on the underlying theory of the origin of the turbulence, or the magnitude of the magnetic torque acting on the fluid. However, a certain degree of universality is expected. Several numerical \cite{Avila2012,Lohse2014num} and experimental \cite{Vernet2022, Paoletti2012, Lathrop1992, Huisman_2012} studies have predicted the existence of \textcolor{black}{a nondissipative  regime} for AM transport at a sufficiently large Reynolds number. Since astrophysical systems exhibit huge Reynolds numbers~\cite{Charbonneau2013}, astrophysical flows are expected to follow such an asymptotic regime independent of any molecular diffusion, which can be a very helpful guide towards building scaling laws and theories for the turbulent transport of angular momentum in astrophysics. \\

\textcolor{black}{A recent numerical study has partially addressed these various questions by modeling a radiative stellar layer as a differentially rotating, electrically conducting, spherical Couette flow subject to stable stratification~\cite{Petitdemange2023} . It has been shown that, as expected, Tayler instability can generate complex flow motions vigorous enough to amplify a magnetic dynamo of high magnitude. In this case, the dynamo is  subcritical, and the associated magnetic torque is strong enough to produce an efficient diffusionless transport of angular momentum and slow down the inner part of the star (see next section for more details). It was also noted that during this transition, the flow bifurcates into a highly fluctuating state. Using a numerical setup identical to that of \cite{Petitdemange2023}, the present article aims to study this bifurcation in more detail, focusing on the nature of the velocity fluctuations and the exact conditions under which this transition is observed.}

\textcolor{black}{We show that a magnetic instability generates small-scale turbulence. This turbulence then reamplifies the large-scale magnetic field via a mean-field dynamo that ultimately sustains the turbulence. In other words, we observe a subcritical bifurcation to a state characterized by turbulence, strong magnetism, and efficient angular momentum transport in a flow that is linearly stable without a magnetic field}. A low-dimensional nonlinear dynamical system is proposed as a comprehensive model that aligns with this mechanism and successfully explains our DNS. Remarkably, our model unveils a simple and relatively universal framework for understanding the subcritical transition to turbulence in astrophysical systems marked by the interplay of a large-scale dynamo, a hydrodynamic transition, and a small-scale magnetic instability.

\section{subcritical turbulence}
\label{section2}

The stellar radiative zone is modeled here as a spherical Couette flow in which an electrically conducting fluid is confined between two spheres with radii $r_i$ and $r_o$  whose rotation rates $\Omega_i=\Omega+\Delta\Omega$  and $\Omega_o=\Omega$ are kept fixed, with the indices $i$ and $o$ representing the inner and outer spheres, respectively. \tck{This imposed global rotation is the simplest way to describe the differential rotation observed in the stellar radiative zone. Our model therefore relies on a fixed shear, whereas it can evolve over time in a real star. In particular, high angular momentum transport leads to a flattening of the rotation profile in stars, which is not reproduced here. An important point of the paper is therefore to elucidate the role of this shear on the flow dynamics. On the other hand, the inner sphere can also be seen as a rough model of the boundary with the convective zone, which is expected to rotate at a different angular velocity \citep{Petitdemange2023}.}

Stable stratification is ensured within the Boussinesq approximation by imposing a constant temperature difference  $\Delta T=T_o-T_i>0$ between the two spheres.  All of our simulations are conducted with the pseudo spectral code PaRoDy \cite{Dormy1998, Aubert2008} using the Shtns library \cite{Shtns2013}. \tck{PaRoDy uses a finite difference scheme in the radial direction and a poloidal-toroidal decomposition on a spherical harmonics basis in the angular directions. Typical resolutions are $336$ points in the radial direction, and $150$ and $60$ in the maximum degree $\ell$ and order $m$ respectively.} It then solves the following MHD Boussinesq equations: 

\begin{eqnarray}
\nonumber \frac{\partial \mathbf{u}}{\partial t} + (\mathbf{u}.\bm{\nabla})\mathbf{u}  = -\frac{\bm{\nabla} \Pi}{\rho_0} - 2  \bm{\Omega} \times \mathbf{u} +  \nu &\Delta \mathbf{u}& + \frac{1}{\rho_0\mu_0}\left(\bm{\nabla}\times\mathbf{B}\right)\times\mathbf{B}+ \alpha_T\Theta g_0(r/r_o)\mathbf{e_r},   \\
\frac{\partial \Theta}{\partial t} + (\mathbf{u}.\bm{\nabla})\Theta &=& \kappa \Delta \Theta  -(\mathbf{u}.\bm{\nabla})T_s,\\
\frac{\partial \mathbf{B}}{\partial t} &=& \bm{\nabla} \times \left(\mathbf{u} \times \mathbf{B} \right) + \eta \bm{\Delta} \mathbf{B}, \\
\nabla. \mathbf{u} &=&0,\,\nabla. \mathbf{B} =0, 
\label{pb_eq}
\end{eqnarray}



where $\mathbf{u}$ , $\mathbf{B}$, and $\Pi$ are, respectively, the fluid velocity, its magnetic field, and total pressure, including the centrifugal force. $\Theta$ is the temperature perturbation, accounting for the variation of density, and  \tck{as in convection studies, $T_s$ is the purely conductive temperature profile, solution of $\Delta T_s=0$}. $\rho_0$ is the fluid density and $\nu,\kappa$, $\eta$, and $\alpha_T$ are, respectively, the kinematic viscosity, thermal diffusivity, magnetic resistivity, and thermal expansion coefficient, \tck{all considered constant both in time and in space}. The gravity field has a linear radial profile $g\propto r$ such that $g_0=g(r_o)$ is the value taken at the outer sphere. \tck{Such a profile is expected for a spherical object whose density is constant, resulting from Gauss’ law for gravity \citep{ChristensenD2008}}. No-slip conditions for the velocity field are imposed on both spherical boundaries. The outer sphere is electrically insulating \textcolor{black}{(both the radial current and the toroidal magnetic field are zero at $r_o$)}, while the inner sphere represents the inner part of the star and therefore corresponds to an electrically conducting boundary (with a conductivity $\sigma$ identical to that of the fluid). \textcolor{black}{ Three different types of initial conditions have been used throughout the article. DNS are either initialized with an infinitesimal seed \tck{($10^{-5}$ in code units)} of fields with symmetric and antisymmetric components \tck{(toroidal $\ell=2,m=0$ and poloidal $\ell=1,m=0$, respectively)}, restarted from a previously computed solution, or initialized with a very strong dipole field. The condition used in practice for each run will be specified below.} This system can be described by six independent, dimensionless control parameters: the Rossby number $Ro=\Delta \Omega/\Omega$ and the Ekman number $Ek=\nu/r_o^2\Omega$, respectively measure large-scale shear  and viscous effects, both compared to global rotation. Stratification is described by the ratio between Brunt–\tck{Väisälä} (or buoyancy) frequency and global rotation $N/\Omega=\sqrt{\alpha_T g_0\Delta T/(r_o-r_i)}/\Omega$. Finally, the magnetic Prandtl number $Pm=\nu/\eta$ and the thermal Prandtl number $Pr=\nu/\kappa$ describe the relative influence of molecular diffusivities. \tck{For simplicity, the aspect ratio is kept fixed to $\chi=r_i/r_o=0.35$, typically corresponding to a three-solar-mass star involving a convective core with a radiative envelope (see, for example, \cite{Charbonneau2013}).} Note that the kinetic and magnetic Reynolds numbers can be easily deduced from these parameters, $Re= \chi Ro/Ek$ and $Rm=RePm$. Similarly, the output magnetic field is measured by the Elsasser number $\Lambda=\frac{\langle B \rangle^2}{\mu\rho\eta\Omega}$, which \textcolor{black}{compares the magnetic force to the Coriolis force, with $\langle\rangle$ denoting a spatial average over the domain}. \textcolor{black}{For the computation of our various bifurcation diagrams and scaling laws}, the reported values are time-averaged over the saturated phase of the stationary state, for at least one ohmic time $t_\eta = (r_o-r_i)^2/\eta$, and taken either locally or spatially averaged where indicated.

Before discussing the dynamo results, we \textcolor{black}{describe here the purely hydrodynamic results obtained in the absence of a magnetic field. At sufficiently low Rossby number, the basic flow state is a stratified axisymmetric flow, composed of a strong azimuthal Couette flow associated with a weak poloidal recirculation. This velocity field is linearly stable until a critical value of the Rossby number, $Ro_h\approx 0.35$ , is reached}, marking the onset of hydrodynamic linear shear instability.
\textcolor{black}{The flow then bifurcates to an $m=1$ nonaxisymmetric pattern, which drifts in the azimuthal direction. It is crucial to emphasize that $Ro_h$ is the onset of a \tck{large-scale} instability and does not correspond to a transition to turbulence: for all values of $Ro$ and $Ek$ reported here and \tck{regardless of} the initial conditions, the flow remains laminar as long as there is no magnetic field, essentially due to the stabilizing effect of stratification.} In other words, we have never observed any subcritical nor supercritical transition to turbulence in the absence of a magnetic field. We do, however, expect such a transition to occur at larger Rossby number, when  shear is large enough to counteract the effect of the stratification (see, for example, \cite{Chang2021}).

\begin{figure}[ht]
    \centering
    \includegraphics[width=0.775\linewidth]{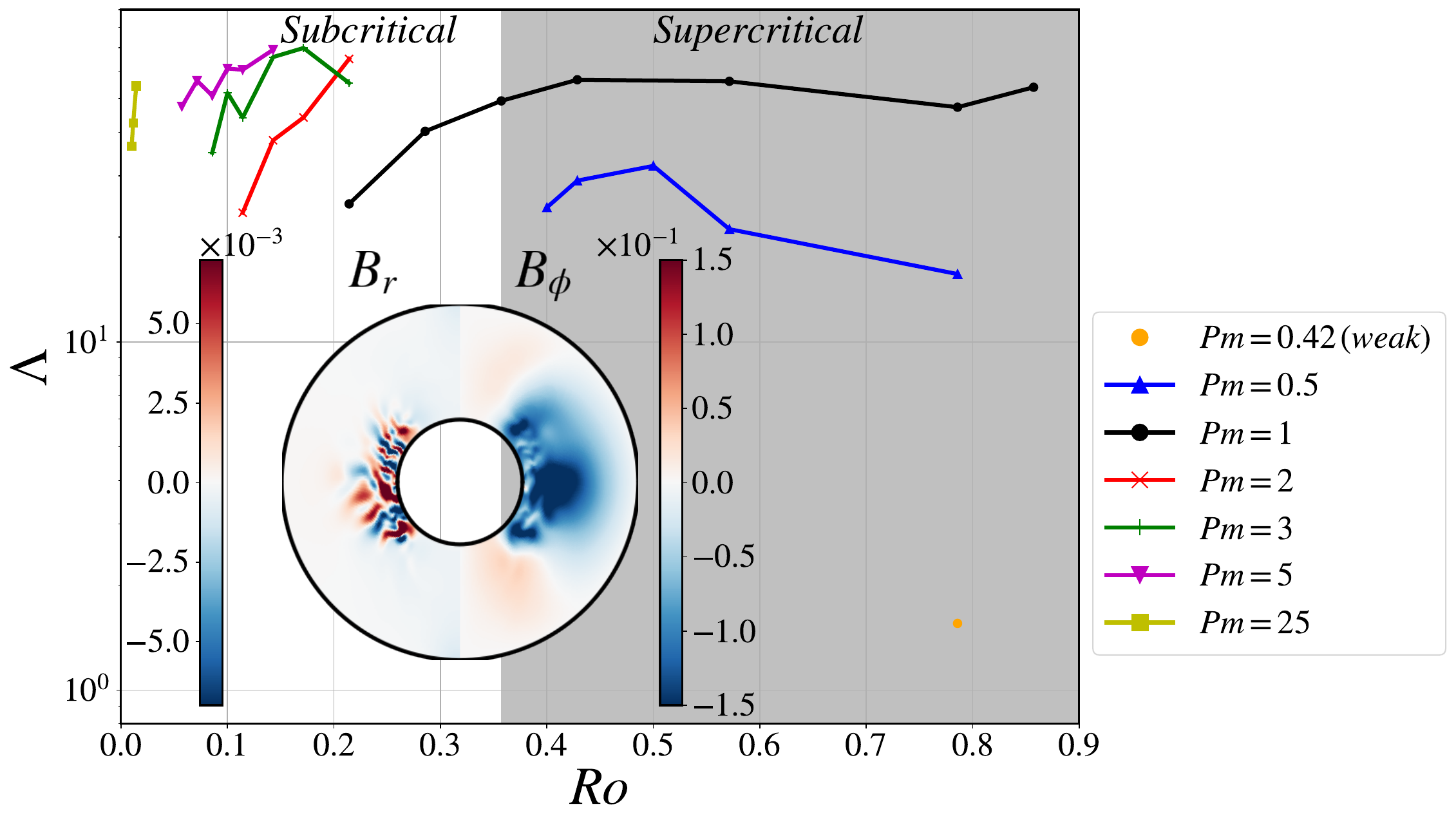}       
    \caption{Bifurcation diagram of Elsasser number vs Rossby number \textcolor{black}{for $Ek=10^{-5}$ and $N/\Omega=1.24$}. \textcolor{black}{Taking advantage of the subcritical nature of the bifurcation, the dynamos at $Ro<Ro_h$ have been obtained either by restarting from the solution obtained at $Pm=1$, or initialized with a dipolar magnetic field strong enough to reach the basin of attraction of the corresponding solution (see text for details)}. Inset: Snapshot of the magnetic field components $B_r$ and $B_\phi$ for the run $Ro=0.14,\,Pm=2$ at $\phi=0$. }
    \label{fig:bif1}
\end{figure}


Let us now switch to the simulations initialized with an infinitesimal magnetic field. The axisymmetric laminar Couette flow below $Ro_h$ is unable to \tck{generate a dynamo from} this initial magnetic field. For $Ro>Ro_h$ however, two types of magnetic field can be amplified and sustained. The first solution is a viscously-dominated magnetic field, which saturates at relatively low value. This state, which we refer to as the weak dynamo, coexists with a laminar velocity field. \textcolor{black}{It is primarily generated by the action of the nonaxisymmetric $m=1$ flow discussed above and appears as a supercritical bifurcation when $Ro$ is increased. It takes the form of a nearly axisymmetric toroidal magnetic field associated with a smaller $m=1$ perturbation. 
The second dynamo branch is radically different, and has been partly discussed in \cite{Petitdemange2023}. The appearance of this solution is strongly correlated with the magnitude of the large-scale toroidal magnetic field $B_\phi$ in the simulation (see Fig.\ref{fig:bif1} for a typical snapshot). Any large value of $B_\phi$, generated by the weak dynamo described above triggers the so-called Tayler instability.  This pinch-type instability is well known to the stellar physics community, as it provides a simple means of destabilizing stably-stratified stellar interiors. It produces nonaxisymmetric perturbations when a toroidal axisymmetric magnetic field becomes sufficiently strong. Linear theory of stratified Couette flows predicts that the Tayler instability of a toroidal magnetic field should generate an $m=1$ perturbation at low Reynolds numbers \tck{\citep{Spruit1999,Kirillov2014}}. In the DNS reported here, when $B_\phi$ reaches the onset of the Tayler instability, it instead produces large, chaotic, multiscale fluctuations in magnetic and velocity fields involving several wave numbers $m$.} Surprisingly, the appearance of these fluctuations coincides with a secondary amplification of the magnetic field leading to the strong-field dynamo solution, \textcolor{black}{following a scenario very similar to that proposed by Spruit \cite{Spruit2002}.}

\begin{figure}[ht]
    \centering
    \includegraphics[width=0.775\linewidth]{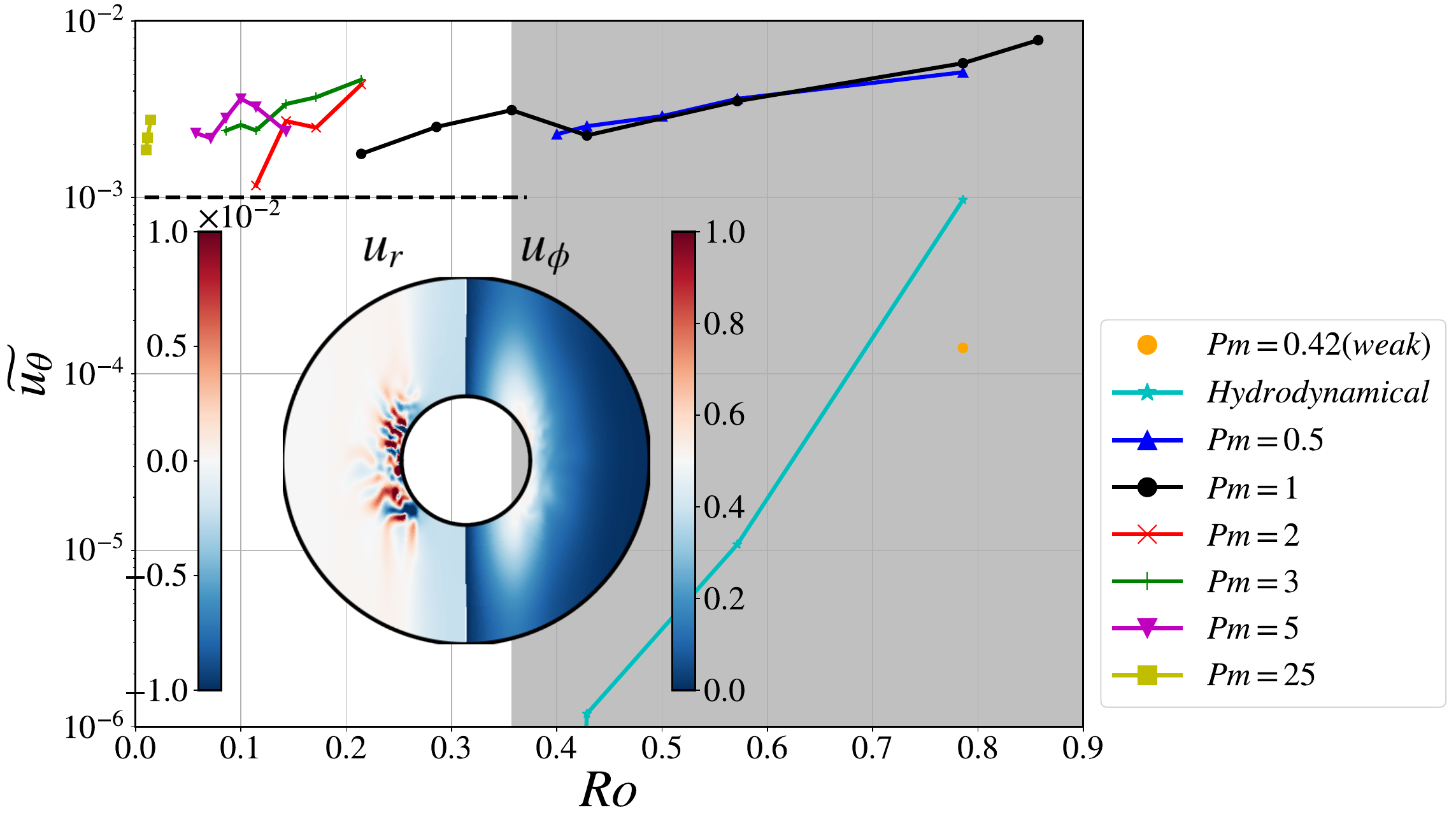}    
    \caption{\tck{Bifurcation of the velocity perturbations as a function of the Rossby number for $Ek=10^{-5}$ and $N/\Omega=1.24.$ For each run, $\displaystyle \widetilde{u_\theta} = \sqrt{\overline{\left<(u_\theta-\overline{u_\theta})^2\right>_t}}$ is measured locally in the equatorial plane where the dynamo is the strongest, at $r=0.55r_o$.} Inset: Snapshot of the velocity field for the run $Ro=0.14,\,Pm=2$ at $\phi=0$.}
    \label{fig:bif2}
\end{figure}

 \textcolor{black}{Following on the single bifurcation reported in \cite{Petitdemange2023},  we have explored here the strong field branch} over a wide range of control parameters, including $Pm$. Figure \ref{fig:bif1} reports a bifurcation diagram of the Elsasser number versus the Rossby number $Ro$ for various simulations. \textcolor{black}{It first shows that the strong-field dynamo exhibits  large values of the Elssaser number ($\Lambda>20$). This illustrates the magnetostrophic force balance achieved on this branch, in which the Lorentz force  is large enough to balance with the Coriolis force}. More importantly, this dynamo solution is highly subcritical, as it can be sustained to arbitrary small differential rotation, even below $Ro_h$. 
 \textcolor{black}{For these simulations performed at $Ro<Ro_h$}, the strong field solution is obtained either by restarting a previous simulation already on the strong field branch \textcolor{black}{(cases $Pm=0.5$, $Pm=1$, $Pm=2$ and $Pm=3$)}, or by initializing the simulation with a strong \textcolor{black}{dipolar} field \textcolor{black}{(cases $Pm=5$ and $Pm=25$)}.  This subcritical dynamo solution can thus be obtained for parameters where neither the weak field dynamo nor the shear instability is present. In the last section, we will nevertheless discuss the importance of the neighbourhood with these linear instabilities.


Figure \ref{fig:bif2} \tck{clarifies} this surprising behavior. It shows the bifurcation of the velocity field fluctuations measured in the dynamo region as a function of $Ro$, for the same set of simulations. \tck{In this figure, these fluctuations are obtained by computing $\widetilde{u_\theta} = \sqrt{\overline{\left<(u_\theta-\overline{u_\theta})^2\right>_t}}$ at $r \approx 0.55r_o$ in the midplane, normalized by the velocity of the inner core $\Delta \Omega r_i$ (In the following, $\overline{X}$ denotes an azimuthal average, while $X^{*}=X-\overline{X}$ and $<>_t$ corresponds to a time average).}

\begin{figure}[ht]
    \centering
    \includegraphics[width=0.75\linewidth]{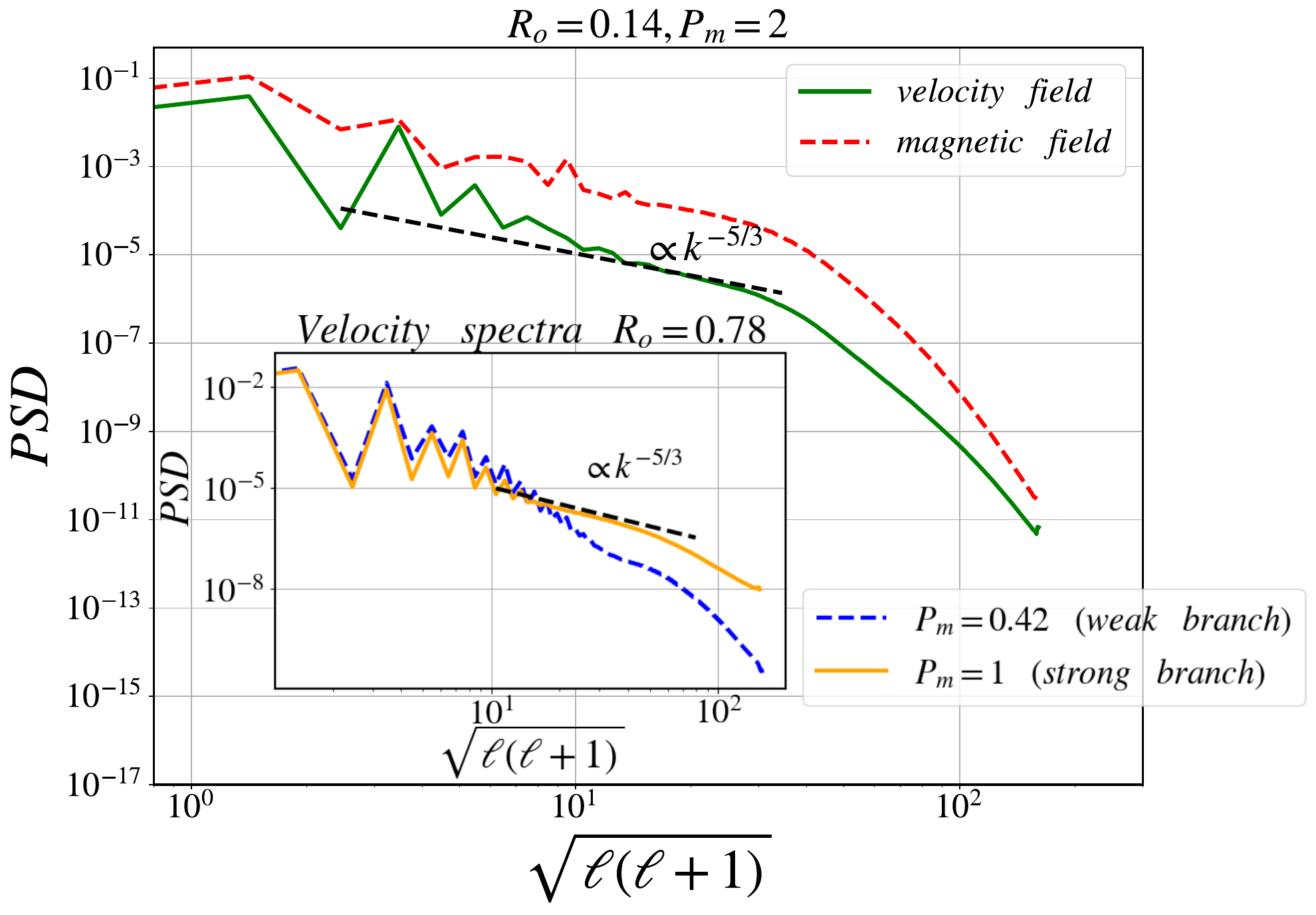}    
    \caption{Power spectrum density for the run $Ro=0.14,\,Pm=2\tck{,\,Ek=10^{-5},\,N/\Omega=1.24}$, for the velocity and magnetic field, computed in the active dynamo region  ($0.4<r/r_o<0.6$) and time-averaged over the saturated phase. \textit{Inset}: PSD for two runs at the same $Ro=0.78\tck{,\,Ek=10^{-5},\,N/\Omega=1.24}$, but for different $Pm$ (strong and weak branches).}
    \label{spectra}
\end{figure}

The cyan curve indicates the nonmagnetic case \textcolor{black}{discussed previously}, where the onset $Ro_h$ of the shear instability distinguishes a linearly stable flow with no perturbations from a linearly unstable flow in which \tck{the amplitude of the} laminar oscillations increases continuously as $Ro$ (or $Re$) is increased. Again, even at the largest $Ro$ (corresponding to $Re=27500$), the strong stratification keeps the flow laminar such that these fluctuations simply correspond to periodic oscillations related to the shear instability. \textcolor{black}{Because the weak dynamo is generated by this shear instability, it displays} similar oscillations (orange point). 
This is in strong contrast with the strong field branch, which displays a high level of fluctuations, reaching up to 5 \% of the mean azimuthal Couette flow $\overline{u_\phi}$. The magnitude of these turbulent fluctuations remains relatively constant over the whole range of explored parameters, although the Reynolds number is varied by almost three orders of magnitude. \textcolor{black}{This highly fluctuating state regenerates the magnetic field on which it feeds, and can thus be easily maintained below the onset $Ro_h$ of the shear instability, at relatively low kinetic Reynolds number.}
\tck{Figure \ref{spectra} shows that these fluctuations on the strong branch exhibit a spectrum compatible with a Kolmogorov K41 energy spectrum generally observed for isotropic three-dimensional MHD turbulence \citep{Biskamp2000}, with an exponent $k^{-5/3}$ for both velocity and magnetic energy. It is, however, difficult to conclude on the exact value of this exponent, and many effects such as anisotropy due to rotation or the presence of a large toroidal field can produce a different phenomenology (see [37], for example, for a recent review of MHD spectra). Nevertheless, note, that this spectrum is considerably less abrupt than that observed on the low-field branch for smaller $Pm$, where the flow is clearly laminar.}


\begin{figure}[ht]
    \centering
    \includegraphics[width=0.49\linewidth]{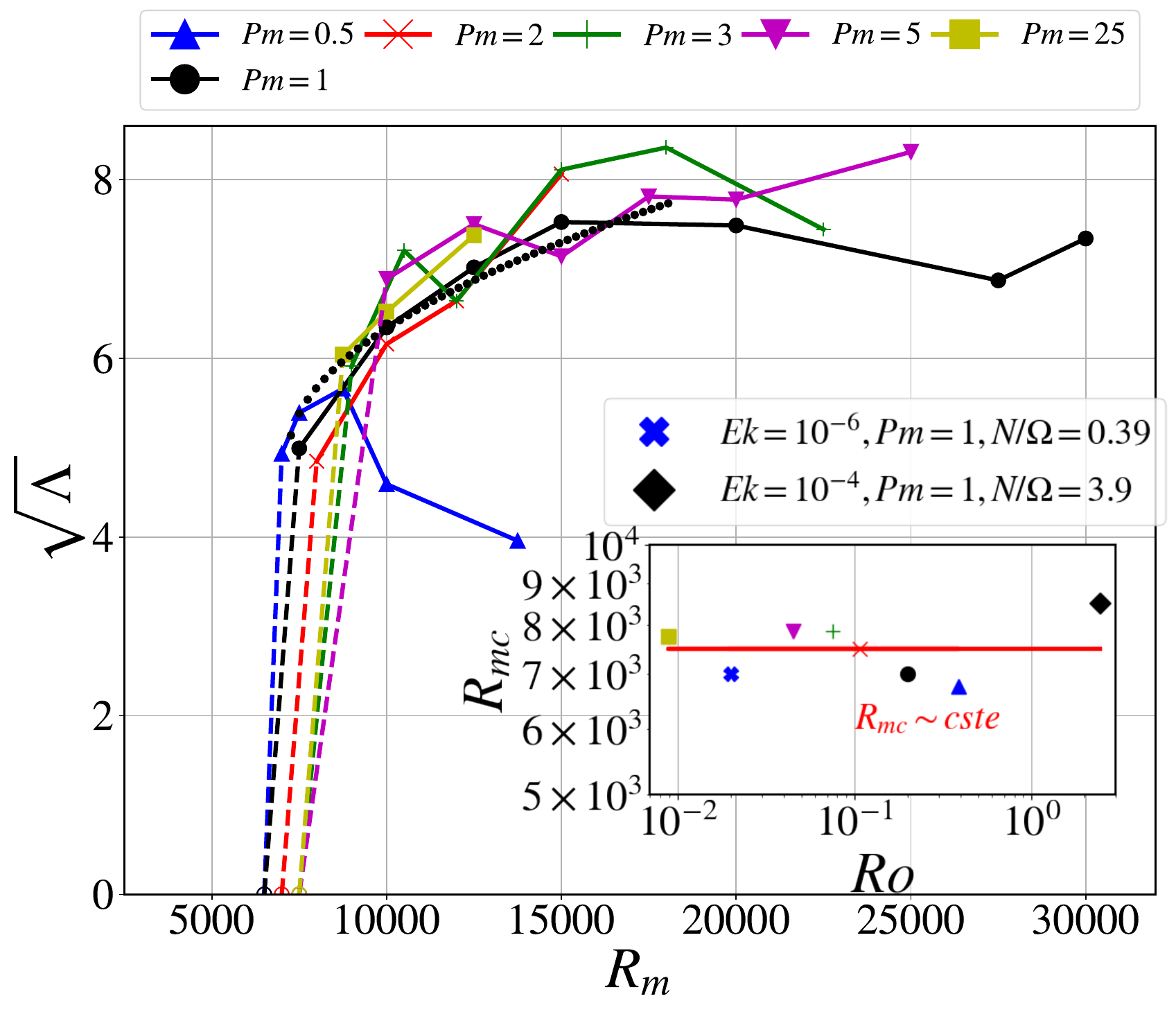}
      \includegraphics[width=0.49\linewidth]{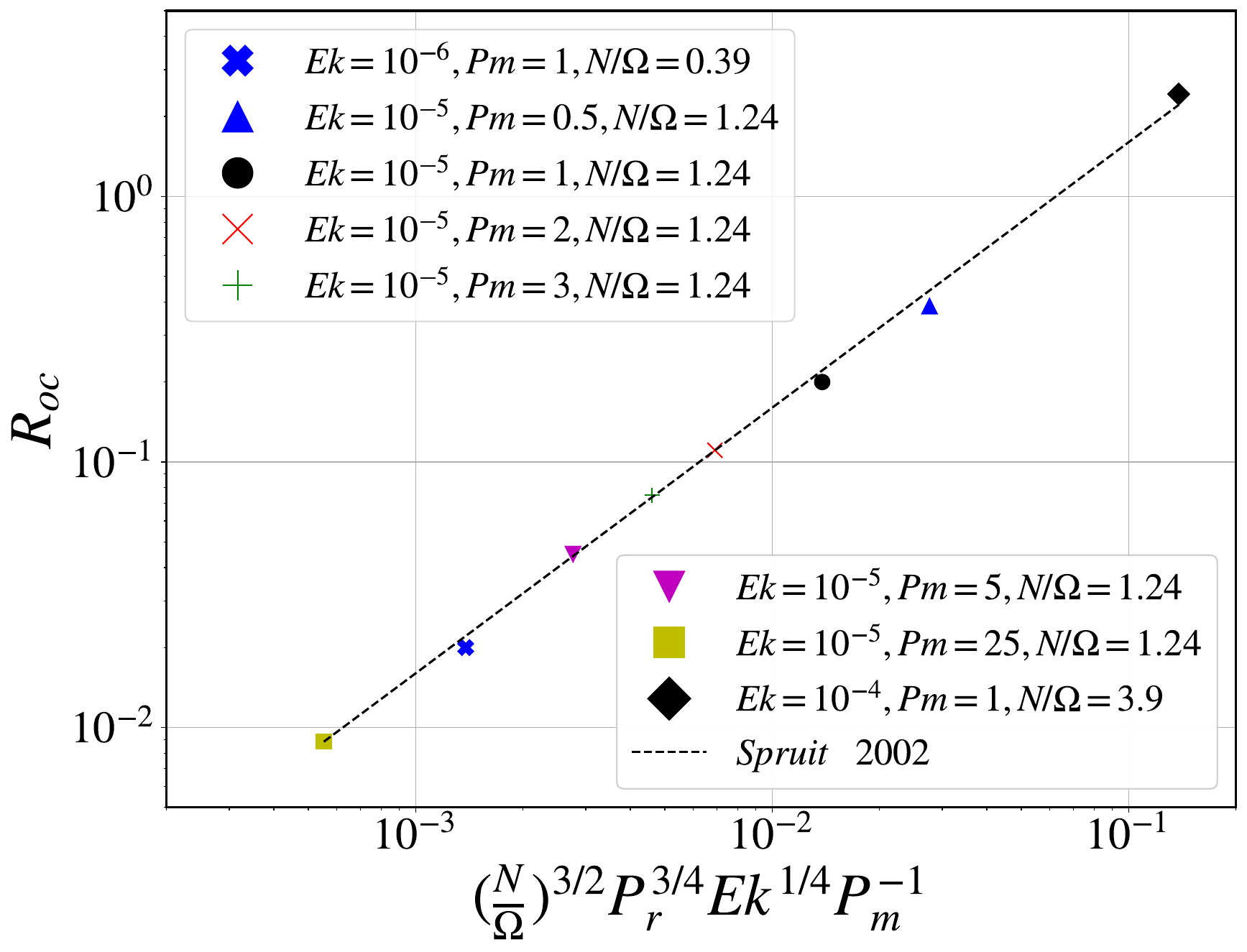}    
    \caption{\textcolor{black}{Left:}  $\sqrt{\Lambda}$ as a function of $Rm$ for different $Pm$. The  normal form of the bifurcation is relatively well fitted by the nonlinear model (dotted line) developed in Sec. \ref{section4}. Inset:  The critical magnetic Reynolds number of the dynamo, computed for each $Pm$, is independent of the kinetic Reynolds number. \textcolor{black}{Right: Critical Rossby number for the dynamo, compared to Spruit's prediction \cite{Spruit2002} (dashed line). }}
    \label{fig:bif3}
\end{figure}

These simulations therefore describe a subcritical transition to turbulence \textcolor{black}{(in terms of the Rossby number $Ro$)} triggered by its interaction with a magnetic dynamo. This subcritical transition is well illustrated by Fig. \ref{fig:bif3} (left), which shows that our data can be rescaled if the saturated Elsasser is plotted as a function of the magnetic Reynolds number $Rm$ \textcolor{black}{rather than $Ro$} (for fixed $Ek$ and $N/\Omega$). \tck{This seems to hold in a large range of $Rm$ for large $Pm$ but is somehow less true for the case $Pm=0.5$, although it also shows some rescaling, but in a reduced range. However, we argue that our numerical setup is relevant for stellar interiors ($Pm<10^{-5}$), as }\textcolor{black}{this rescaling with $Rm$ means} that the generation of the strong field dynamo is not necessarily related to the large values of $Pm$, but rather relies on the generation of a high level of induction. \textcolor{black}{It also means that the magnetic Reynolds number $Rm$ controls how far from the onset $Ro_h$ the subcritical dynamo can be maintained}. Data from the DNS are very well fitted by the theoretical prediction (dashed line) of the model proposed in the last section, and corresponding to the typical normal form of a subcritical bifurcation. This bifurcation point $Rm_c$ is also associated with a critical value of the Elsasser number, below which no dynamo can be observed ($\Lambda_c\sim20$). This clearly reflects the need for the dynamo to be in magnetostrophic equilibrium, and can indeed be seen as the main constraint for the Maxwell stress scaling law recently proposed for such radiative stellar layers \cite{Petitdemange2023}. 
This rescaling also implies that the critical onset of the dynamo $Rm_c\sim 7000$ is independent of the kinetic Reynolds number \textcolor{black}{or the magnetic Prandtl number and is only controlled by the level of stratification and global rotation} (see inset). This is in sharp contrast with usual DNS of turbulent dynamos, which almost always display an increase of the onset with the level of turbulence, followed or not by a saturation to constant $Rm$ at large $Re$ \cite{Schekochihin2004, Ponty2005}. This can be understood as a consequence of the subcritical nature of the transition to turbulence, which necessarily implies that the small-scale velocity fluctuations provide the induction mechanism for the large-scale field. In contrast, the coherence of the large-scale flow is strongly affected by the turbulence, but apart from the provided shear, it is not directly involved in the dynamo mechanism.

 \textcolor{black}{Incidentally, this regime results in turbulent flow at kinetic Reynolds numbers well below the classical values expected for turbulence \tck{in such setups (see \cite{Wicht2014} for the particular case of unstratified spherical Couette flow, or \cite{Petitdemange2023b} for the stratified case)}. More importantly, this turbulence can be maintained orders of magnitude below the primary linear onset of the flow}. It simply requires the magnetic Reynolds number to be kept above some critical value $Rm_c$ so the strong field dynamo is maintained. At $Pm=25$ for instance, the critical Rossby number for dynamo action is $Ro_c=8\times 10^{-3}$, which is $45$ times smaller than the critical onset $Ro_h$ for linear shear instability. \textcolor{black}{This illustrates Spruit's minimum level of differential rotation required to sustain the dynamo  \cite{Spruit2002}. Indeed, Eq. (27) in  \cite{Spruit2002} predicts a minimum shear rate required to obtain dynamo action, which, translated into our dimensionless parameters, reads $Ro_c = (N/\Omega)^{3/2}(Pr^3Ek)^{1/4}/Pm$. As shown in Fig. \ref{fig:bif3} (right), this prediction is extremely well satisfied in our simulations.}  
 
A final comment can be made on the existence of this critical magnetic Reynolds number for the dynamo:   for numerical convenience, the global differential rotation between the two spheres is imposed in our setup, while the angular velocity profile in real \tck{stellar} radiative zones can be extremely flat \tck{(see for instance \cite{Couvidat2003} for the sun or \cite{Aerts2019} for bigger stars)}. As the reported dynamo is very efficient to transport angular momentum, it is expected to flatten the rotation profile \cite{Petitdemange2023}. The fact that dynamo can be obtained for such low values of differential rotation is therefore an important finding of our simulations, because we can expect angular momentum transport to be sustained until nearly solid body rotation is achieved, provided that the magnetic Reynolds number remains above some critical value ($Rm_c\sim 7000$ in the present simulations).


\section{Angular momentum transport}

 \begin{figure}[!htb]
    \centering
    \includegraphics[width=0.875\linewidth]{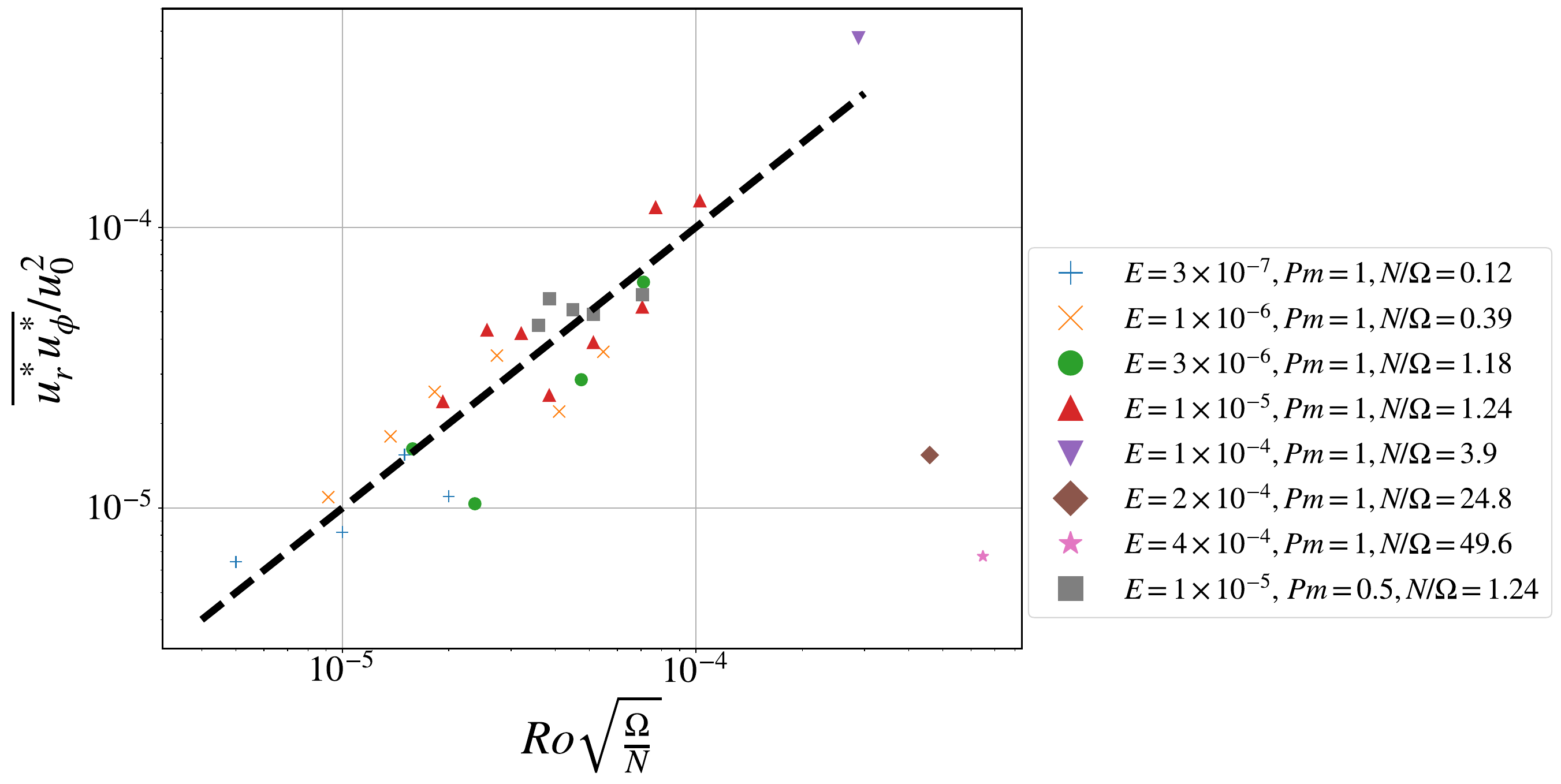}    
    \caption{Scaling law for the dimensionless Reynolds stress $\overline{u_r^*u_\phi^*}/u_0^2$. This Reynolds stress is measured in the dynamo region, for $\pi/4<\theta<3\pi/4$ and $0.45<r/r_0<0.55$. Here we show simulations only for $Pm\le 1$, as they correspond to simulations with the highest Reynolds number, relevant for astrophysical applications (see discussion).}
    \label{fig:scaling1}
\end{figure}

During the last decades, several predictions have been made for the angular momentum (AM) transport in stellar radiative interiors, depending on the mechanism involved in the destabilization of the stably-stratified shear flow. Internal waves, GSF instability, or the Tayler-Spruit dynamo can all contribute significantly to the AM transport, and generally lead to several different theoretical scaling laws. 
In the classical picture of the Tayler-Spruit mechanism \cite{Spruit2002}, the dynamo arises when the axisymmetric toroidal magnetic field $\overline{B_\phi}$ is large enough to trigger the Tayler instability \cite{Tayler1973}, for which the linear theory predicts a nonaxisymmetric $m=1$ perturbation in the flow \cite{Acheson1978}. This perturbed motion then re-amplifies the initial magnetic field and closes the loop for an exponential amplification of the toroidal field. This naive picture was initially criticized  by Zahn \cite{Zahn2007}, because of the impossibility of regenerating an axisymmetric toroidal magnetic field using perturbations due to the Tayler instability, which are essentially nonaxisymmetric. 
Our results show a quite different story and resolve this controversy without rejecting the Tayler-Spruit picture: at large values of the magnetic Reynolds number, the Tayler instability generates fluctuations of the magnetic field $b^*$ much more complex than the simple $m=1$ perturbation predicted by the linear theory. The corresponding chaotic structure, involving a large range of different length scales, is well illustrated by the snapshot shown in Fig.\ref{fig:bif1}. In the magnetostrophic regime, these magnetic fluctuations induce chaotic fluid motions $u^*$ through the Lorentz force, which produce a turbulent spectrum characterized by many wave numbers $m$ and an effective Tayler length scale located in the middle of the inertial range (see Fig.\ref{spectra} or snapshot in Fig.\ref{fig:bif2}).  This small-scale turbulence then generates a mean electromotive force ${\cal E}=\langle u^* \times b^* \rangle$ which  re-amplifies the  axisymmetric toroidal field and closes the loop. \\

This picture requires large kinetic and/or magnetic Reynolds numbers and local generation of small-scale turbulence independent of boundary conditions, a situation encountered in most astrophysical flows. In this case, as described in the introduction, AM transport is expected to depend only weakly on the molecular diffusion. In the simplest possible model of a radiative stellar interior in which molecular coefficients are ignored, the Reynolds stress $\overline{u_r^*u_\phi^*}$ then depends on only the typical velocity scale $u_0$, the typical integral length scale $r$, the buoyancy frequency $N$, and the rotation rate $\Omega$. Dimensional analysis yields a simple prediction of the form $\overline{u_r^*u_\phi^*}/u_0^2=Ro^a(\Omega/N)^b$, and most theories proposed in the literature fit within this prediction with $a$ and $b$ depending on the destabilization mechanism.

Figure \ref{fig:scaling1} shows that our turbulent simulations are well described by such a scaling law:

\begin{equation}
\frac{\overline{u_r^*u_\phi^*}}{u_0^2} \sim Ro \sqrt{\frac{\Omega}{N}},
\label{reynolds}
\end{equation}

where \textcolor{black}{$u_0\approx 5.10^{-2}(r\Delta\Omega)$ is the bulk velocity in the dynamo region and is only a fraction of the large-scale shear due to the presence of Ekman boundary layers~\cite{Petitdemange2023}}.
It can  be rewritten as $\rho\overline{u_r^*u_\phi^*}\sim \frac{\rho u_0^3}{\Omega r}\sqrt{\Omega/N}$, an expression very close to the one proposed for AM transport by internal waves at the base of the solar convection zone \cite{Press1981, Zahn1996}. Following Press and Zahn, we interpret $\rho u_0^3$ as the mechanical energy flux of the large-scale shear flow, part of which is drained off to the kinetic energy flux of small-scale fluctuations. The magnitude of this conversion is related to some impedance matching depending on the $\Omega/N$ ratio, so that strongly stratified flows display a small-scale energy flux much less than  $\rho u_0^3$. In this respect, it should be noted that this scaling law could be modified for very large values of $N/\Omega$ relevant to stellar applications (see the Discussion for more details).

\section{Nonlinear model}
\label{section4}

\begin{figure}
    \centering
    \includegraphics[width=0.6\linewidth]{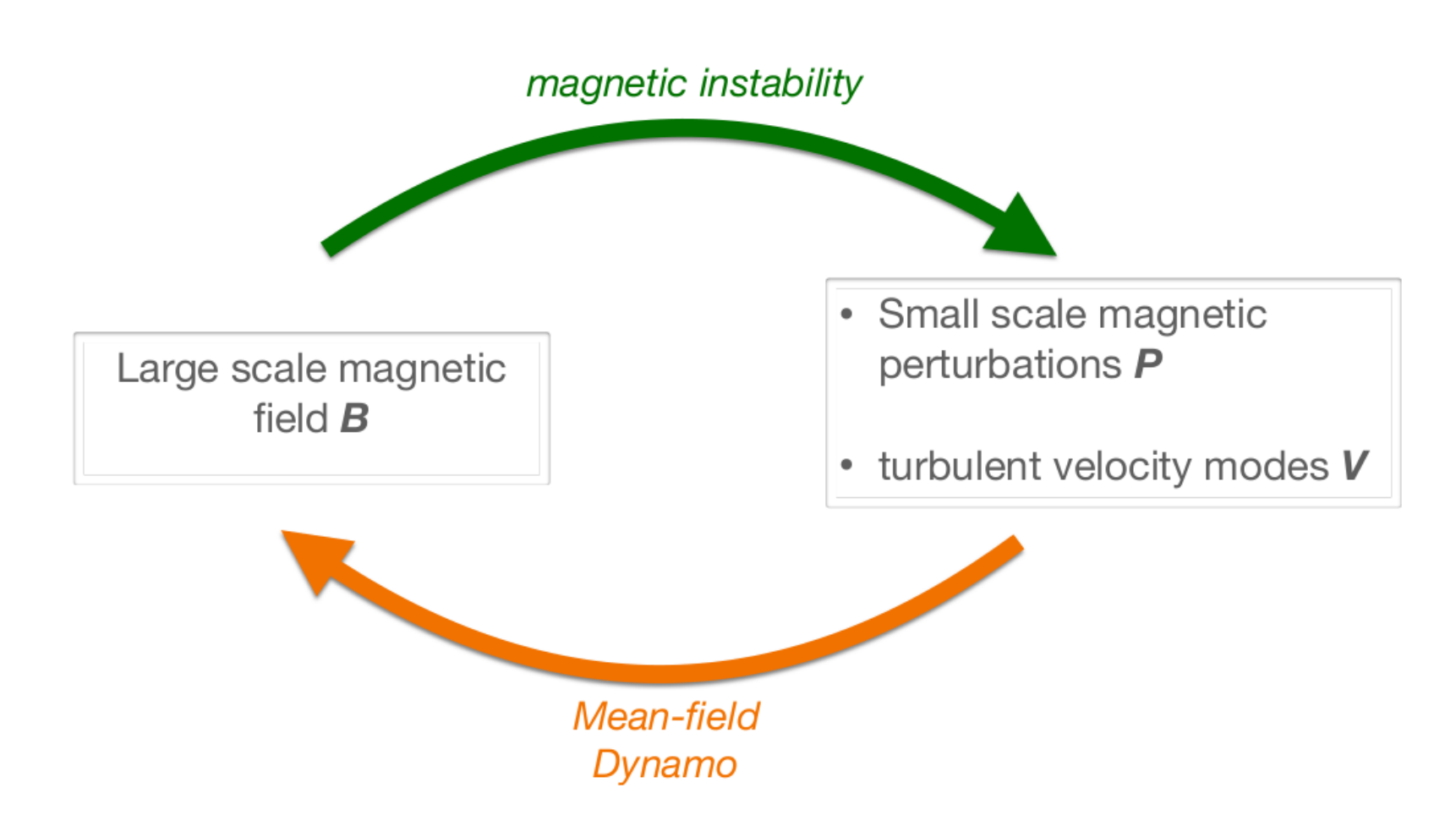}
    \caption{\tck{Diagram illustrating} the mechanism described by the model in Eqs. (\ref{modelB})-(\ref{modelV}): a large-scale magnetic field $B$ is destabilized through some linear MHD instability, generating small-scale perturbations. At large $Rm,Re$, these perturbations are chaotic and drive turbulent velocity fluctuations, which then regenerate the initial large-scale field through a mean-field dynamo. }
    \label{fig:schema_model}
\end{figure}

The scenario described here joins a long list of mechanisms proposed to explain the origin of turbulence and magnetic fields in astrophysics, and \textcolor{black}{relying on a finite amplitude magnetic field}. These include the generation of turbulence in accretion disks by the magneto-rotational instability \cite{Balbus1991}, the generation of a strong magnetic field by nonlinear processes in radiative stellar layers \cite{Fuller2019, Stefani2019, Ji2023}, or instabilities in convective stellar layers driven by magnetic buoyancy for instance \cite{Cline2003}. Despite significant differences between these models, the amplification loop that sustains both magnetic field and turbulence can \textcolor{black}{often}  be summarized by the \tck{diagram} in Fig.\ref{fig:schema_model}. In this picture, the large-scale magnetic field undergoes a magnetic instability (MRI, Tayler instability, magnetic buoyancy, etc.) producing small-scale velocity fluctuations which then regenerate the initial large-scale field through dynamo action. 
\textcolor{black}{Keeping in mind that many astrophysical dynamos and turbulent flows do not rely on a subcritical mechanism,} this process provides a simple explanation for the joint amplification of magnetic field and velocity fluctuations.\\

Our simulations of the fluctuating Tayler-Spruit dynamo are consistent with this picture, but also show that the transition is highly subcritical and systematically associated with a mean-field dynamo sustained by turbulent motions.  The aim of this final section is to propose a simple nonlinear model capturing this scenario, and general enough to be applied to other systems. We start by recalling the classical equations generally used to describe a mean-field dynamo \cite{Moffatt2019}: 
\begin{eqnarray}
\partial_t B_T &=& s(B_P.\nabla) \frac{U}{s} + \nabla\times (\alpha B_P)+\eta \nabla^2 B_T, \label{mean_field_eq_toro}\\
\partial_t A   &=& \alpha B_T + \eta\nabla^2A,
\label{mean_field_eq}
\end{eqnarray}

where $B_T$ (respectively $B_P=\nabla\times A e_\phi$ ) describes the  toroidal (resp. poloidal) component of the large-scale field, $U$ is  the large-scale azimuthal axisymmetric shear flow \tck{and $s$ is the cylindrical radius $s=r\sin\theta$}. Several types of magnetic amplification can be produced depending on the terms involved in the process. The first term of the right-hand side of Eq. (\ref{mean_field_eq_toro}) is the most important: this $\omega$ effect describes the amplification of the toroidal magnetic field by the winding of the poloidal field lines by the large-scale shear.  On the other hand, the poloidal magnetic field can be regenerated from the toroidal field due to the $\alpha$ term in Eq. (\ref{mean_field_eq}), which represents the  electromotive force due to the average contributions of small-scale flow motions. In some cases, the toroidal field can also be amplified through the alpha effect represented by the second term of the right-hand side of Eq. (\ref{mean_field_eq_toro}).

\begin{figure}
    \centering
    \includegraphics[width=0.6\linewidth]{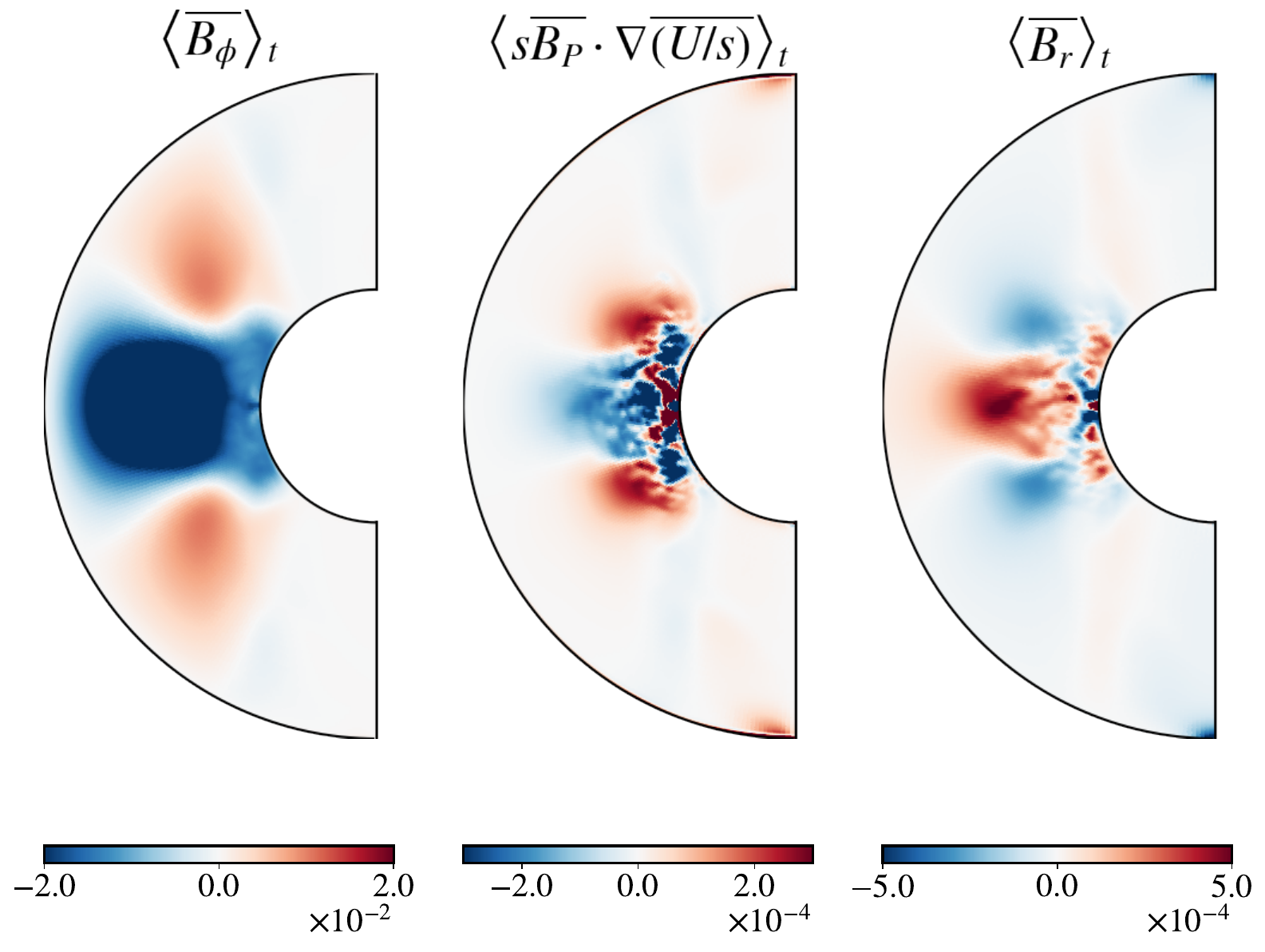}
    \caption{\textit{From left to right:} (a) map of $\overline{B_\phi}$, averaged in time and in the azimuthal direction. (b) Same, for the $\omega$-effect, $\overline{B_P}\cdot \nabla\overline{U}$. (c) Same for  $\overline{B_r}$. All the maps are computed on the saturated phase for $R_o=0.78,\, Pm=1 \tck{,\, Ek=10^{-5},\, N/\Omega=1.24}$ .}
    \label{fig:alpha_omega}
\end{figure}

Figure \ref{fig:alpha_omega} shows the time-averaged maps of $\overline{B_\phi}$, \tck{$\overline{sB_P}\cdot \nabla\overline{(U/s)}$} and $\overline{B_r}$, and indicates that the first two are strongly correlated, i.e., that the dynamics of our stellar layers relies on an $\omega$-effect, similarly to classical  $\alpha-\omega$ dynamos \cite{Schrinner2012}.

But in the presence of a subcritical transition, turbulence is in fact directly constrained by the magnetic field, so that $\alpha$ depends on the magnetic modes and Eqs.  (\ref{mean_field_eq_toro})-(\ref{mean_field_eq}) become strongly nonlinear.
Rather than using a model based on $\alpha$ parametrization, a general model for the subcritical transition to turbulence in magnetized flows requires explicit treatment of the small-scale velocity and magnetic modes as dynamic variables. In the following, we therefore construct a nonlinear model for the evolution of both magnetic and velocity modes involved in the dynamics. \textcolor{black}{Interestingly, the dynamics reported in our DNS can be very well captured by reducing the governing equations to a nonlinear dynamical system involving only a few of these modes.}

Let us first decompose the fields into axisymmetric and nonaxisymmetric parts: 

\begin{eqnarray}
{\bf B}({\bf r},t)&=&B(t){\bf B_{axi}}({\bf r})+\left(\sum_m P_m(t){\bf b_m}({\bf r}) + c.c. \right),\\
{\bf u}({\bf r},t)&=&\sum_m V_m(t){\bf u_m}({\bf r}) + c.c.,
\end{eqnarray}

where ${\bf B_{axi}}$ is now the \textcolor{black}{spatial structure of the axisymmetric  component of the magnetic field} and $B$ its (real) amplitude (without distinction between the poloidal or toroidal component). Its nonaxisymmetric component is described by a complete spectrum of complex amplitudes $P_m(t)$ corresponding to different azimuthal wave numbers $m$, and c.c. designates the complex conjugate of the previous expression. The phase of $P_m$ then describes the angle of the corresponding pattern in the equatorial plane.  A similar decomposition is used for the nonaxisymmetric velocity field. \textcolor{black}{Only a small number of these modes are expected to be involved in the dynamics. In principle, to identify these modes, one could perform a weakly nonlinear perturbative expansion of the complete equations \cite{Guckenheimer1983}. Instead, we follow here the usual approach for pattern-forming systems \cite{GFD1991, van1994amplitude}, in which the amplitude equations are determined by symmetry arguments and by the general properties of the underlying physical problem.}

\textcolor{black}{ In the Tayler-Spruit theory, the dynamo pattern is a combination of $m=1$ nonaxisymmetric magnetic and velocity fields (the linear eigenmodes of the Tayler instability) coupled to a large-scale toroidal field. Our model must therefore, at the very least, reproduce this interaction between a nonaxisymmetric magnetic mode $P$, a nonaxisymmetric velocity mode $V$ (corresponding to the same wave number $m$ as $P$) and the large-scale magnetic field $B$ such that $\partial_tB=f(V,B,P)$, $\partial_tP=g(V,B,P)$, and $\partial_tV=h(V,B,P)$.} 
This does not mean that the other modes are unimportant or necessarily stable, but rather that they can be adiabatically eliminated without profoundly altering the dynamics of the large-scale fields which display slow dynamics. \textcolor{black}{This choice also provides the minimal ingredients needed to capture the mechanism illustrated in Fig.\ref{fig:schema_model}. Note that although the $m=1$ mode is the most natural, the derivation below is in fact independent of the wave number chosen, so the model applies to any value of $m$.}

\textcolor{black}{Next, as our model aims to represent a weakly nonlinear expansion of the complete equations, $f$, $g$ and $h$ can be written as a power series expansion around the origin $B=P=V=0$. Most of the possible terms of the series, however, are precluded by the symmetries of the problem}. So, using the rotational invariance $V\rightarrow Ve^{im\chi}$, $P\rightarrow Pe^{im\chi}$, the fundamental symmetry $(B,P)\rightarrow (-B,-P)$ of the induction equation, and the requirement that the velocity mode be subject to a pitchfork bifurcation when no magnetic field is present, only a few terms remain.  By limiting the expansion to cubic terms at most, we finally obtain a simple system of equations of amplitude :

%

\begin{eqnarray}
\dot{B}&=&(-\mu  + c_1|V|^2)B -\alpha_1B^3,
\label{modelB}\\
\dot{P}&=&(-\nu+ \beta_2 B^2)P + c_2VB -\beta_1|P|^2P ,
\label{modelP}\\
\dot{V}&=&\lambda V +\gamma BP -c_3|V|^2V. 
\label{modelV}
\end{eqnarray}

\textcolor{black}{Based solely on the symmetries and general properties of the original problem, these three coupled equations represent the simplest model for capturing the nonlinear dynamics between a large-scale magnetic dynamo and a smaller-scale MHD instability. \tck{Except for the signs of the parameters and the choice of saturation terms (suggested by DNS), no other assumptions were made to obtain this model.} As we show below, however, it perfectly reproduces the phenomenology described in Fig.\ref{fig:schema_model}. and in the DNS.}

The first equation describes the generation of a large-scale field $B$ by the small-scale velocity field $V$. In the absence of this nonaxisymmetric flow ($V=0$), the large-scale magnetic field $B$ must decay, hence the (minus) term $-\mu$. Note that this first equation for $B$ can also be rigorously derived from the mean-field equations (\ref{mean_field_eq_toro})-(\ref{mean_field_eq}): the parameter $\mu$ describes the combined action of shear and magnetic diffusivity on the large-scale field, providing a linear amplification (or attenuation) of this field. \textcolor{black}{In particular, this means that large-scale flow dynamics are not explicitly described, but rather incorporated into the model parameters.} On the other hand, a mean-field dynamo generally involves an $\alpha$ term related to the kinetic helicity $u^*.\nabla\times u^*$. The $V^2B$ term can therefore be considered as a crude model of the electromotive force produced by the small-scale turbulence. \textcolor{black}{Note that a saturation term $P^2B$ , allowed by the symmetries, has been discarded such that the equation is only saturated by the classical and simpler $B^3$ term}.

The second equation describes the dynamics of the nonaxisymmetric magnetic mode. In the case of the Tayler-Spruit dynamo, this may be one of the unstable modes of the Tayler instability. More generally, $P$ represents a small-scale mode generated by any instability of the large-scale magnetic field. It is therefore important to note that our set of equations is valid for any wave number $m$, provided that $P$ and $V$ correspond to similar wave numbers. This mode $P$ is zero if $B=0$, but can be linearly amplified \textcolor{black}{by the term $B^2P$ above a critical value of the mean field, if $B^2>|\nu|/\beta$}. This explicitly describes the Tayler instability, the MRI instability or any linear magnetic instability involved in the cycle illustrated in Fig.\ref{fig:schema_model}. \textcolor{black}{The  term $VB$ is the lowest-order term compatible with the symmetries and  describing magnetic induction by the nonaxisymmetric velocity}. In particular, in the case where $m$ represents rotational symmetry, this last term guarantees that an exponential growth of axisymmetric $B$ is necessarily associated with a growth of nonaxisymmetric $P$, as predicted by Cowling's theorem for nearly axisymmetric flows. More generally, the two terms $VB$ and $V^2B$ describe the part of the cycle in Fig.\ref{fig:schema_model} corresponding to the generation of the mean-field dynamo.

The subcritical nature of the \textcolor{black}{bifurcation observed on the DNS leads us to explicitly write} an equation for the velocity mode. This equation describes the linear instability of this small-scale mode and can be related to any hydrodynamic instability (at $B=0$) of the flow, typically the \textcolor{black}{shear instability observed in our simulations, such that $\lambda\propto (Ro-Ro_h)$}. The $PB$ term is the lowest order term representing the feedback of the quadratic Lorentz force on the flow \textcolor{black}{(we ignore higher order magnetic terms $B^2V$ and $P^2V$)}. This term describes how perturbations of the magnetic field $P$ can induce turbulent fluctuations $V$, which is essential for capturing a subcritical transition to turbulence. \textcolor{black}{Finally, note that negative signs are used for the terms $B^3$, $|V|^2V$ and $|P|^2P$, so these cubic terms provide field saturation, as usual. }

The equations for $B$ and $V$ with $P=0$ describe the close competition between a hydrodynamic instability (involving only $V$) and a dynamo instability (involving only $B$). Although the nonlinear terms are different, this system is therefore relatively close to the normal form for a codimension-2 bifurcation point, which has been studied in detail by many authors (e.g. in \cite{Guckenheimer1983}). 
\textcolor{black}{This type of model is well known for describing subcritical transitions due to the competition between two instabilities occurring at the same point (in this case, $\lambda=\mu=0$), and has even been proposed in the context of subcritical dynamo bifurcations \cite{krstulovic2011axial,marcotte2016dynamo}}. The additional equation for $P$ complicates the analysis of the dynamical system, which deviates considerably from these previous models. \textcolor{black}{Finally, note that although $P$ and $V$ represent complex amplitudes, the parameters used are real, and no terms involving complex conjugates are present. This means that the dynamic of the phase of the modes is ignored, as it does not control the subcritical nature of the bifurcation. The new set of equations then} perfectly captures the dynamics observed in DNS: it models the additional competing instability of the large-scale magnetic field generating small-scale magnetic perturbations $P$.

\begin{figure}[h]
    \includegraphics[width=0.9\linewidth]{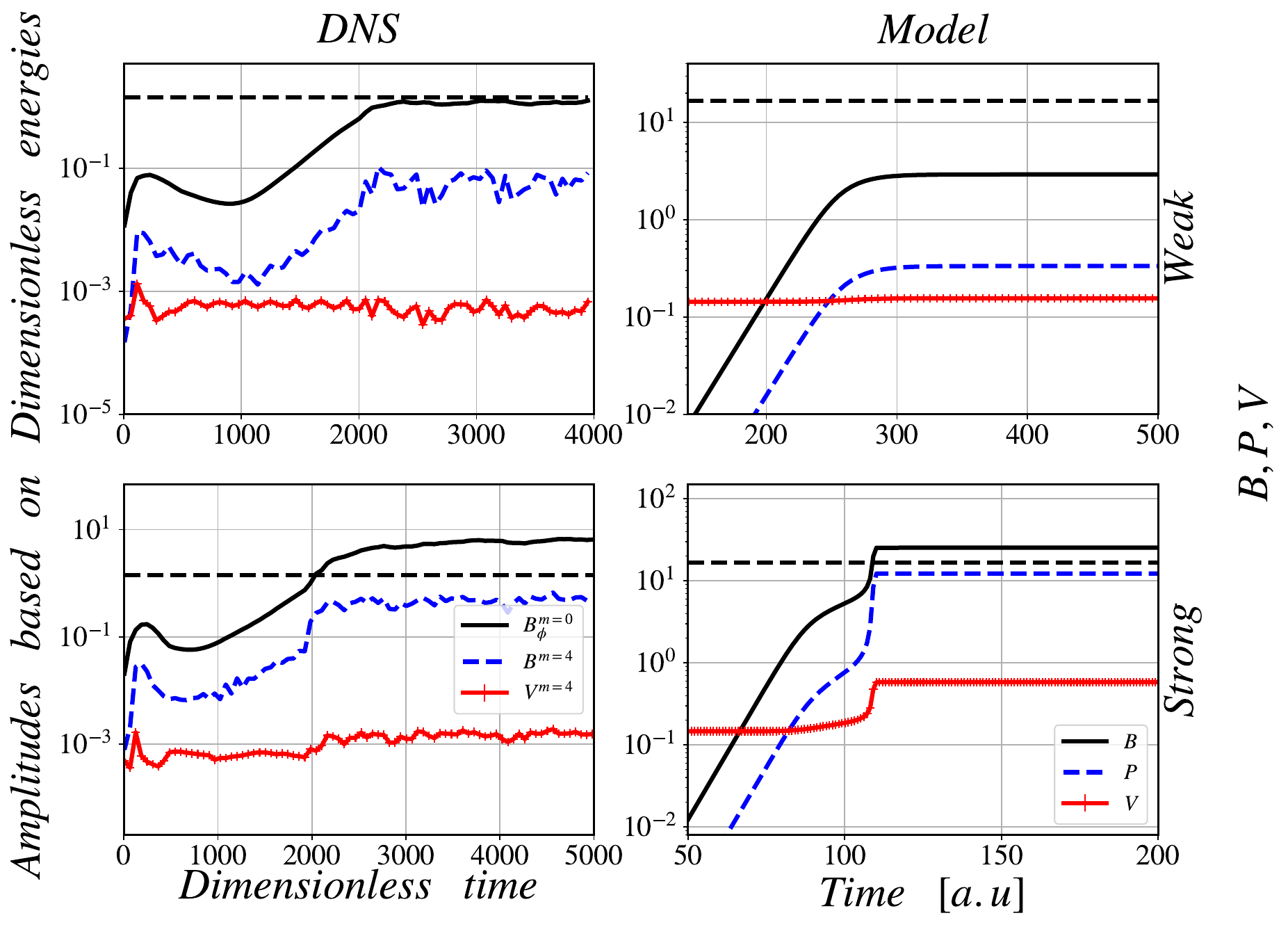}
    \caption{Timeseries from our DNS \tck{at $Ek=10^{-5}, N/\Omega=1.24, Ro=0.78$} (left) compared to the ones from the model derived here (right) for the weak (top, \tck{$Pm=0.42$}) and the strong (bottom, \tck{$Pm=1$}) dynamo branches. The horizontal dashed lines correspond to Tayler instability threshold, \tck{computed locally as in \cite{Petitdemange2023}}. For DNS, we plot $\sqrt{\Lambda}$ for the magnetic field, and $\sqrt{E_{kin}}$ for the velocity field, where $E_{kin}$ is obtained by averaging the full kinetic energy on the whole volume on the saturated phase. Equations (\ref{modelB})-(\ref{modelV}) of the nonlinear model are integrated using the Odeint module of Python, with $\mu=2$, $\alpha_1=0.05$, $c_1=100$, $\nu=7$, $\beta_1=0.1$, $\beta_2=0.025$, $c_2=5$, $c_3=100$, $\gamma=0.06$, using $\lambda=2.05$ (weak) or $\lambda=2.15$ (strong).}
    \label{fig:model1}
\end{figure}

Figure \ref{fig:model1} compares the time series obtained from the model with those observed in DNS. 
\textcolor{black}{Parameter values are shown in the legend, but it should be noted that similar behaviors can be obtained for a wide range of parameter values. On the other hand, different behaviors can also be observed for other parameters. Only a thorough perturbative expansion of the initial governing equations can provide the exact parameter values corresponding to the case simulated in DNS. But generally speaking,} the model predicts two different magnetic dynamos depending on the parameters. In the top right-hand corner, a first dynamo ($B,P\ne 0$) grows for sufficiently large $V$, but saturates at relatively low values, so that the magnitude of the small-scale mode is only slightly modified. This first solution is in very good agreement with the weak branch dynamo observed in DNS (upper left corner). The strong branch is also reproduced by the model, as shown in Fig.\ref{fig:model1}, bottom right corner: the time series first starts with the weak field dynamo, until $B^2$ reaches the critical value $\nu/\beta$, producing a secondary amplification up to very large values, thus profoundly modifying the saturation of the velocity mode. Here again, the agreement with DNS is very good: this secondary amplification is also observed as soon as the toroidal field reaches the threshold of the Tayler instability (lower left corner). Note also that during secondary amplification of $B$, the value of the small-scale mode $V$ is increased. Our model therefore predicts an increase in the amplitude of $V$ as the magnetic instability sets in. In our DNS, only velocity modes with azimuthal wave numbers $m>4$ exhibit such behavior, with  $m<4$ modes decreasing as the large-scale field is re-amplified. The exact wave number involved in these dynamics may vary from simulation to simulation (see \cite{Petitdemange2023b}). However, small-scale structures with larger wave numbers are expected to adiabatically follow the dynamics of the leading mode, so $V$ may hopefully provides an adequate tracer of the dynamics of small-scale turbulence.  \textcolor{black}{Note that the parameters in Fig.\ref{fig:model1} have been chosen so that the reduced model has mode amplitudes comparable to those of DNS, which facilitates comparison between the two systems.}

\begin{figure}[h]
    \includegraphics[width=0.7\linewidth]{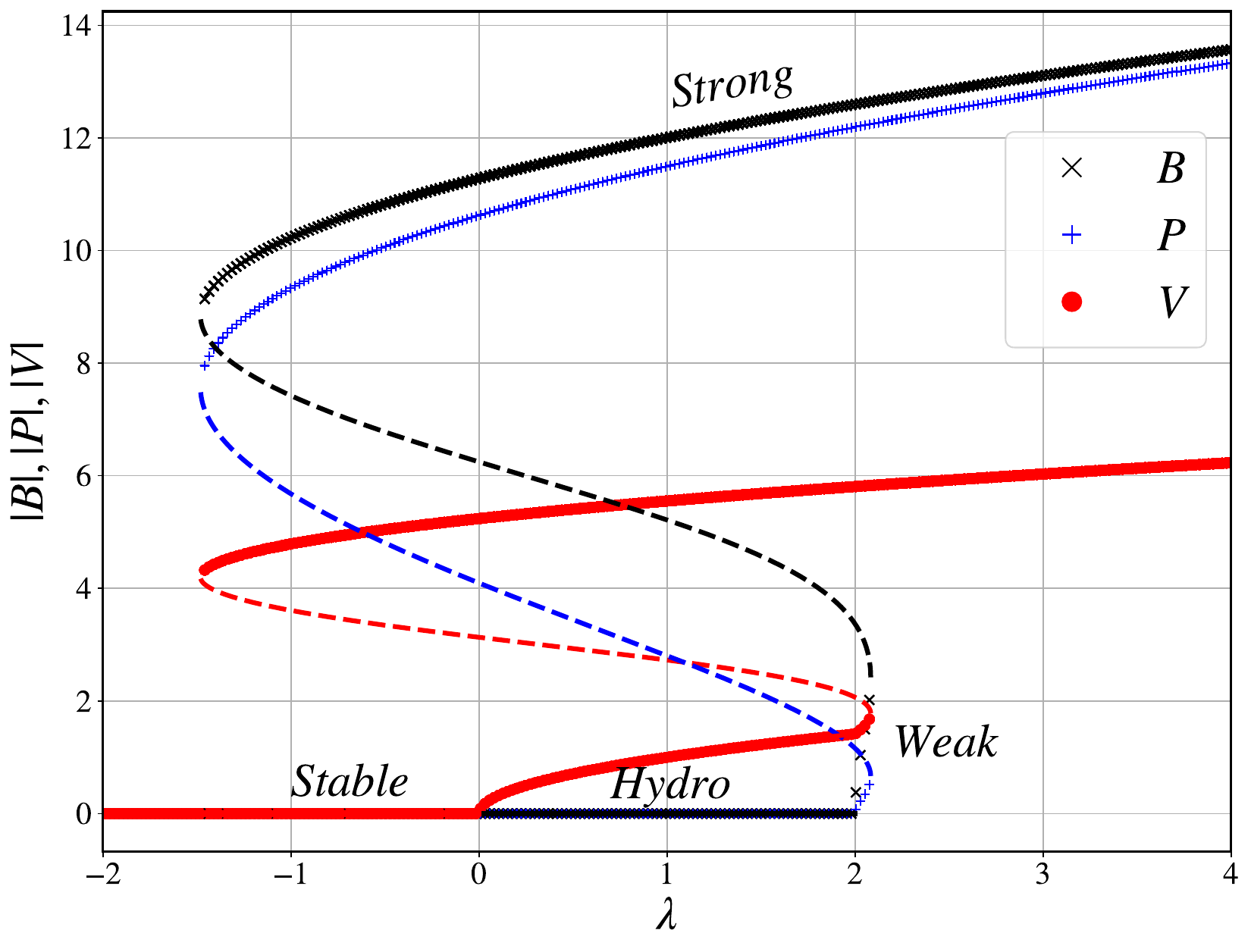}
    \caption{Bifurcation diagram for the amplitude of each mode as $\lambda$ varies. The parameters values are $\mu=2$, $\alpha_1=0.2$, $c_1=1$, $\nu=7$, $\beta_1=0.1$, $\beta_2=0.1$, $c_2=1$, $c_3=1$, $\gamma=1.2$. The points correspond to stable solutions that have been obtained by direct simulations of the system (\ref{modelB})-(\ref{modelV}), while the dashed lines correspond to unstable solutions that have been computed analytically.}
    \label{fig:model2}
\end{figure}

Figure \ref{fig:model2} shows the bifurcation diagram obtained in the model, which reproduces the bistability between a weak laminar dynamo on one hand, and a strong large-scale dynamo associated with small-scale velocity fluctuations on the other hand. As expected, the weak dynamo can only be generated above the hydrodynamic threshold $\lambda\propto (Ro-Ro_h)>0$ and displays a supercritical pitchfork bifurcation. In contrast, the strong branch is subcritical: it is associated with much larger values of the magnetic field $B$ and the small-scale fluctuations $V$  can be maintained  well below the linear threshold $\lambda<0$. Thus, the model reproduces fairly well turbulence and magnetic field generation in a linearly stable region, far from any linear shear instability.
For comparison, we have superimposed in Fig.\ref{fig:bif3} (black dots) the solution of our dynamical system by rescaling the amplitudes of the modes and using numerical values indicated in Fig.\ref{fig:model2} with $\lambda\propto(Rm-Rm_c)$. 
\textcolor{black}{
As shown in Fig.\ref{fig:model2}, the subcritical nature of the strong-field branch also produces a large gap in the value of the axisymmetric field $B$ between the subcritical solution, characterized by a very strong field, and the supercritical solution associated with a much lower magnitude. This leads to the generation of a forbidden gap very similar to the so-called magnetic desert that has been observed for intermediate-mass stars \cite{lignieres2013dichotomy}.
}

\section{Discussion and conclusion}

By focusing solely on the angular momentum transported by the mean magnetic field, the nature of the turbulence generated in stellar interiors is often overlooked. The results presented here illustrate how magnetism and turbulence are deeply intercorrelated, and clarify in particular how turbulence \tck{can} arise in stably stratified radiative stellar layers. \tck{We can thus see that the paradox usually mentionned for the Tayler-Spruit model \tck{\citep{Zahn2007}}, of a nonaxisymmetric instability regenerating a $m=0$ poloidal field}, naturally disappears if dynamo is associated with a transition to fully developed turbulence: \tck{turbulent fluctuations generate an electromotive force whose axisymmetric component is large enough to regenerate the original axisymmetric poloidal magnetic field, as predicted by classical mean-field theory}. Our simple nonlinear model shows that this necessarily implies a subcritical transition to turbulence, which is assured as long as $Rm$ and $Re$ are sufficiently large.

Taking this turbulence into account also modifies the predictions for AM transport, as the Reynolds stress due to turbulent velocity fluctuations produces an additional transport given by the relation (\ref{reynolds}). A comparison between the turbulent transport described here and that achieved by the large-scale magnetic field, $B_rB_\phi/(\mu\rho)\sim \sqrt{r}(u_0\Omega)^{3/2}/N$ (see \cite{Petitdemange2023}), shows that the ratio between the former and the latter scales as $\sqrt{Ro^3\frac{N}{\Omega}}$. In agreement with previous studies based on different arguments \cite{Fuller2019}, this suggests that transport due to turbulence should, in the vast majority of cases, be smaller than that due to the mean Maxwell stress. In the case of rapid rotators, however, fluctuations can make a significant difference. 
\textcolor{black}{For some subgiants and red giants as the one reported in \cite{Aerts2019}, the Rossby number is around $10$ and $N/\Omega\approx10^3$, so the angular momentum transport due to turbulence could be one order of magnitude larger that the one produced by the Maxwell stress, assuming that our value $u_0\approx 5.10^{-2}r\Delta\Omega$ inferred from the data can be adequately transposed to local stellar rotation rates.}
However, this simple extrapolation to astrophysical regimes must be treated with some caution. Our highly stratified simulations, $N/\Omega=25$ and $N/\Omega=50$, display a Maxwell stress relatively identical to the other simulations \cite{Petitdemange2023} but have a much lower Reynolds stress and do not align with the scaling law (\ref{reynolds}). These simulations are indeed characterized by very large values of $Q=Pr\left(\frac{N}{\Omega}\right)^2$, known to control the geometry of the flow \cite{Philidet2020, Garaud2002}. With $Q=62.5$ and $Q=250$, they correspond to a shellular rotation profile very different from the other runs, which have $Q\sim 1$. A full test of our scaling law would therefore require simulations at large $N/\Omega$ and small $Pr$, a situation extremely difficult to achieve with current numerical ressources.


 
\textcolor{black}{
As discussed above, the scenario observed in our simulations shares many similarities with other theories discussed in the literature. Perhaps the most striking similarity concerns the MRI-driven dynamo (see \cite{Guseva2017} for instance), in which a large-scale dynamo field is supposed to develop from a nonaxisymmetric $m=1$ instability of the azimuthal magnetic field, the so-called azimuthal MRI (AMRI).  In fact, it has been shown that in some cases it is relatively difficult to distinguish Tayler instability (TI) from AMRI \cite{Rudiger2018, Kirillov2014}, which would make the latter an equally valid interpretation of our DNS. In particular, since our simulations display a typical shear flow $U_\phi$ of the same order as the Alfvén velocity $v_A=B\phi/\sqrt{\mu\rho}$, the dominant energies that the instability taps into are comparable.
However, several arguments suggest that the Tayler-Spruit dynamo remains the most convincing explanation: first, nonaxisymmetric perturbations occur systematically near the inner sphere, where the magnetic field increases outwards, unlike AMRI, which is unstable in regions where the azimuthal magnetic field decreases outwards. The critical value of $B_\phi$ for the appearance of this instability is also in agreement with Tayler's predictions. Another strong argument in favor of TI is the onset of our Fig.\ref{fig:bif3}, which shows that the critical magnetic Reynolds number of the dynamo is independent of $Pm$, which aligns perfectly with Spruit's prediction (as discussed in Sec. \ref{section2}), again supporting a Tayler-Spruit dynamo.
That said, the proximity between the two theories clearly suggests that such an AMRI dynamo could be at work in other simulations, or even within real stars. With this in mind, it is important to conclude this paper by discussing the relevance of our nonlinear model to different mechanisms. Relatively universal, it seems that our low-dimensional model can be applied to both mechanisms, as long as they remain based on the $B,P,V$ interaction. In fact, only the value of the coefficients in the Eqs. (\ref{modelB})-(\ref{modelV}) reflects the physical details of the underlying system. A full exploration of the model's dynamics as a function of these coefficients  is naturally beyond the scope of this paper, but may help in the future to distinguish the behavior predicted by these two different theories and describes the transition from one regime to the other.}

Note also that since their derivation is mainly based on general symmetry arguments,  Eqs. (\ref{modelB})-(\ref{modelV}) could just as well apply to any instability-breaking translational symmetry \textcolor{black}{which make them applicable to other mechanisms not necessarily based on the breaking of the rotational symmetry. But the small number of assumptions made to derive the model and} the fact that the subcritical transition is observed over a wide range of these parameters suggests a certain universality in the scenario described here: any instability of a large-scale magnetic field occurring in a linearly stable flow at large $Rm$ and $Re$ may be subject to a subcritical transition to turbulence. 
\textcolor{black}{The excellent agreement between our turbulent DNS and the low-dimensional \{$B,V,P$\} dynamics confirms that the latter can drive complex turbulent systems. It helps to understand the nonlinearities involved in  the universal dynamics  that could be displayed by a number of other physical systems.}
 In particular, the dynamics of the model is related to the coexistence of a hydrodynamic instability and a magnetic instability, interacting through dynamo action. Interestingly, many astrophysical situations have been identified as involving this feature : for instance, it has been proposed that the dynamics of convective stars rely on the coexistence of a Kelvin-Helmoltz hydrodynamic instability and magnetic buoyancy instability \cite{Cline2003}. In protoplanetary disks, the concept of dynamo-MRI also fits this description \cite{Balbus1998}, with a dynamo generated by perturbations driven by the magneto-rotational instability in the vicinity of the (stable) centrifugal instability. A more recent example is the close interaction between saturation of the Tayler magnetic instability and the shear instability in radiative stars \cite{Ji2023}.  All these different examples could well exhibit dynamics similar to those predicted by our model, underlining the relatively universal character of a subcritical transition to magnetized turbulence. Finally, it is worth noting that this model has three modes with quadratic and cubic nonlinearities. It can thus exhibit very rich dynamics, ranging from periodic solutions to chaotic behavior that would be interesting to compare with recent astrophysical observations (\cite{Mitra2005} for instance).


\section{Acknowledgements}
This work was supported by funding from the French program
'JCJC' managed by Agence Nationale de la Recherche (Grant No. ANR 19-CE30-0025-01), from CEFLPRA contract 6104-1 and the Institut Universitaire de France. This study used the HPC resources of MesoPSL financed by
the Région île-de-France and the project EquipMeso (reference
ANR-10-EQPX-29-01) of the program Investissements
d'Avenir, supervised by the Agence Nationale pour la Recherche.

\bibliography{biblio} 

\end{document}